\documentstyle[aps,preprint]{revtex}
\begin{document}
\tightenlines

\title{Moment Ratios for Absorbing-State Phase Transitions}
\author{Ronald Dickman$^{\dagger,a,b}$ and
Jafferson Kamphorst Leal da Silva$^{\ddagger,c}$}
\address{$^{\dagger}$Departamento de F\'\i sica,
Universidade Federal de Santa Catarina, Campus Universit\'ario - Trindade,
CEP 88040-900, Florian\'opolis - SC, Brasil \\ and \\
$^{\ddagger}$Departmento de F\'\i sica,
Universidade Federal de Minas Gerais, Caixa Postal 702,
CEP 30161-970, Belo Horizonte - MG, Brasil}
\date{\today}
\maketitle
\begin{abstract}
We determine the first through fourth moments of the order
parameter, and various ratios,
for several one- and two-dimensional
models with absorbing-state phase transitions.  We perform a
detailed analysis of the system-size dependence of these ratios,
and confirm that they are indeed universal for three models ---
the contact process, the A model, and the pair contact process ---
belonging to the directed percolation universality class.
Our studies also yield a refined estimate for the critical
point of the pair contact process.
\vspace {0.3truecm}

\end{abstract}
\vspace{1.0truecm}

\noindent $^a${\small electronic address: dickman@fisica.ufsc.br } \\
$^b${\small On leave of absence from: Department of Physics and Astronomy,
Herbert H. Lehman College, City University of New York,
Bronx, NY, 10468-1589.} \\
$^c${\small electronic address: jaff@fisica.ufmg.br }
 
\newpage

\section{Introduction}

Testing the universality hypothesis, and establishing
the universality class of the critical points exhibited
by various spin models and field theories
has been a major preoccupation of statistical physics
for some time \cite{cardy}.  In this ongoing project, the utility of
studying a variety of universal quantities beyond
the critical exponents, such as
amplitude ratios and scaling functions, is well established.
In particular, obtaining values for cumulant ratios
of the order parameter has proven an efficient method
for identifying the universality class \cite{binder}.  In many cases,
Binder's ``reduced fourth cumulant,"
$q_4 \equiv 1 - \langle \rho^4 \rangle /3\langle \rho^2 \rangle$
(here $\rho$ represents the order parameter and the brackets denote a
stationary average), is remarkably insensitive to finite-size effects,
permitting one to determine the universality class with a
modest computational effort \cite{binder87}.

While critical phenomena in nonequilibrium systems have been
under intensive study for a good twenty years, and controversies
regarding the universality classes
have frequently arisen, moment ratios have not, to our
knowledge, been applied in this study.  Our aim in this paper
is to extend the method to nonequilibrium
models.  We focus on one- and two-dimensional models exhibiting an
{\em absorbing-state phase transition} (i.e., between an
active phase and one admitting no escape or evolution)\cite{privbook};
all belong to the universality class of directed
percolation \cite{torre,kinzel}.  In the following section we define
the models.  Sections III and IV present our results for one- and
two-dimensional models, respectively.  Section V contains a brief summary.

\section{Models}

The models considered here are all examples of
{\em interacting particle systems}: Markov processes
whose state space is a set of particle configurations
on a lattice, with transitions involving local processes
of particle creation or annihilation \cite{liggett,konno,marro}.
In the {\em contact process} (CP) \cite{teharris},
each site of the hypercubic lattice ${\cal Z}^d$ 
is either vacant or occupied by a particle.
Particles are created at vacant sites at rate $\lambda n /2d$, where $n$ is
the number of occupied nearest-neighbors, and
are annihilated at unit rate, independent of the surrounding
configuration. The order parameter is the particle
density $\rho$; it vanishes in the vacuum state, which is absorbing.
As $\lambda$ is increased beyond $\lambda_c$,
there is a continuous phase transition from the vacuum 
to an active steady state; for $\lambda > \lambda_c$, 
$\rho \sim (\lambda - \lambda_c)^{\beta}$.
The A model was devised as a simplified description of
surface catalysis \cite{rdmb87}; in the present notation
we may define it as a generalized CP in which the creation
rate (at a vacant site) is $\lambda$ for $n > 0$, i.e., as long as the
site has at least one occupied neighbor.  Since creation
occurs more readily in the A model than in the CP, the critical
creation rate $\lambda_c$ is smaller
(1.7417 for the A model versus 3.2978 for the CP, in one dimension),
but the two models (and indeed the whole family of generalized
contact processes), share the same critical behavior \cite{gcp},
namely, that of directed percolation (DP).

The {\em pair contact process} (PCP) has a somewhat more complicated
dynamics \cite{pcp1,pcp2}; transitions 
only occur in the
presence of a nearest-neighbor pair of particles.  Specifically,
in one-dimension, if sites i and i+1 are both occupied, then
the particles at these sites mutually annihilate at rate $p \leq 1$,
while at rate $1-p$ they create a new particle at either i-1 or i+2
(chosen at random, each with probability 1/2), provided the
chosen site is vacant.  The PCP exhibits an active phase for $p < p_c$;
above this value the system falls into one of an infinite number
of absorbing configurations. (Any arrangement of particles devoid
of nearest-neighbor pairs is absorbing.)  While the presence of
infinitely many absorbing configurations is associated with
nonuniversal dynamics, the static critical
behavior, which concerns us here, again falls in the DP
class \cite{pcp2,munoz}.

In equilibrium spin systems 
the magnetization (or its projection along a chosen direction),
takes positive and negative values with equal likelihood, but
in the present case the order parameter is non-negative.
This means that we can study odd as well as even moments
of the order parameter.   In the following sections we report
the first through fourth moments, $m_1,...m_4$, where
$m_n \equiv \langle \rho^n \rangle$,
and the ratios $m_4/m_2$,
$K_2/m_1^2 $ and
$K_4/K_2^2$, where $K_n$ is the $n$-th cumulant of the
order parameter, in particular,
\begin{equation}
K_2 = m_2 - m_1^2
\label{K2}
\end{equation}
and
\begin{equation}
K_4 = m_4 - 4m_3 m_1
 - 3 m_2^2 + 12 m_2 m_1^2 - 6 m_1^4 
\label{K4}
\end{equation}

The argument for universality amongst moment or cumulant
ratios follows the same lines as in equilibrium \cite{binder87}.
We first note that at the critical point,
$m_1 \simeq A L^{-\beta/\nu_{\perp}}$, where $\beta$ is the
order parameter critical exponent, $\nu_{\perp}$
is the critical exponent governing the correlation
length, $L$ is the linear extent of the system,
and $A$ is a nonuniversal constant.
From finite-size scaling, we have that at the critical point,
the probability density for $\rho$ satisfies
$P(\rho,L;\lambda_c) = P(\rho/m_1)
\simeq (L^{\beta/\nu_{\perp}}/A)
{\cal P}(\rho L^{\beta/\nu_{\perp}}/A)$, where
${\cal P}$ is a universal scaling function.
It follows that
$m_n \simeq  A^n L^{-n\beta/\nu_{\perp}} I_n$,
where
\begin{equation}
I_n = \int_0^{\infty} u^n {\cal P}(u) du
\end{equation}
is model- and size-independent.  Thus ratios
of the form $m_n/(m_r^i m_s^j)$ are universal for
$ir + js = n$, as are cumulant
ratios such as $K_4/K_2^2$.

\section{Results For One-Dimensional Models}

\subsection{The Contact Process}

We studied the one-dimensional CP on periodic lattices of $L=20$, 40,...,
320
sites, at the critical point,
$\lambda = \lambda_c = 3.297848$ \cite{ijrd93}.
The sample sizes range from $2 \times 10^6$ for $L=20$ trials to
$2 \times 10^5$ for $L=320$;
the maximum duration of each sample realization
extends from from 400 to $16000$.
All trials start from a fully occupied lattice; we
extract moments from the results for the surviving
sample, when it has relaxed to the quasi-stationary state.

The CP density moments are listed in Table \ref{cpmom};
cumulant ratios are listed in Table \ref{cprat}.
From the latter we see that $K_2/m_1^2 $ converges more
rapidly than $K_4/K_2^2$:
between $L=160$ and $320$ the value of the former
changes by about 0.8\%, the latter by about 2\%.
The ratio $m_4 /m_2^2 $
is remarkably stable, changing by only about 0.3\%
between the two largest $L$ values.  Similar
trends are seen for the A model and the PCP (see below).

We also determined moment ratios for off-critical
values of $\lambda$; 
in Fig. 1 we
plot $K_2/m_1^2 $ versus $\lambda$.
As $L$ increases, the point at which
the curves intersect rapidly approaches $\lambda_c$.
(We find that
the crossing points $\lambda_{cr} (L,2L) $
for each pair of successive $L$ values
follows $\lambda_{cr} (L,2L) \simeq \lambda_{cr} (\infty) + const./L^2$,
with $\lambda_{cr} (\infty) = 3.29785(8) $,
consistent with the best available estimate
for $\lambda_c$.)

\subsection{The A Model}

We studied the one-dimensional A model
on periodic lattices of $L=20$, 40,..., 320
sites, at the critical point,
$\lambda = \lambda_c = 1.74173$ \cite{ijrd93}.
Sample sizes range from $2 \times 10^6$
for $L=20$ to $2 \times 10^5$ for $L=320$;
the maximum duration of each sample realization 
extends from from 200 ($L=20$) to $2 \times 10^4$ ($L=320$).
The density moments are listed in Table \ref{ammom};  cumulant ratios are
listed in Table \ref{amrat}.  The latter are seen to take
values very similar to those found for the CP.

\subsection{Pair Contact Process}

In this model, the order parameter $\rho$ is the density
of nearest-neighbor particle {\em pairs}.
We studied the one-dimensional PCP on periodic
lattices of $L=20$, 40,..., 640
sites, for $p = 0.07708$, 0.07709, 0.07710, and 0.07712.  
The sample sizes range from $10^7$ for $L=20$ to
$2 \times 10^5$ for $L=640$;
the maximum duration of each realization
extends from from 200 to $10^4$.
Previous work yielded $p_c = 0.0771(1)$ \cite{pcp1,pcp2}, but we
decided to try to sharpen this
estimate.  To this end we analyzed the
scaling of the lifetime, $\tau (p,L)$, defined as follows.
Starting with all sites occupied, the probability that a trial survives
(remains active) until time $t$ decays $\sim exp[-t/\tau(p,L)]$.
At the critical point,
the lifetime has a power-law dependence on system size,
$\tau (p_c,L) \sim L^{\nu_{||}/\nu_{\perp}}$, while
for $p \neq p_c$ deviations from a power law are seen.
In Fig. 2 we plot
$\ln t^* \equiv \ln \tau (p_c,L) - (\nu_{||}/\nu_{\perp}) \ln L$
versus $\ln L$, using
$\nu_{||}/\nu_{\perp} = 1.5822$,
the value for the DP class in 1+1 dimensions \cite{ijrd93}.
It is evident that the data for $p=0.07709$ are consistent with a power
law,
while those for the other $p$ values are not,
allowing us to conclude that $p_c = 0.077090(5)$.
The order parameter moments --- for $p = p_c = 0.07709$ --- are
listed in Table \ref{pcpmom};  cumulant ratios are
given in Table \ref{pcprat}.

\begin{table}
\caption{\sf Density moments for the critical CP.
Numbers in parentheses denote statistical
uncertainties in the last figure(s).}
\begin{center}
\begin{tabular}{|r|l|l|l|l|} 
$L$    &    $m_1$ &       $ m_2 $    &    $m_3$     &  $m_4$ \\
\hline\hline
$20$  &   0.47427(7)   &  0.26000(6) &   0.15654(6) &  0.10075(5)    \\
$40$  &   0.39754(5)   &  0.18399(4) &   0.09392(3) &  0.05144(3)    \\
$80$  &   0.33367(8)   &  0.13011(5) &   0.05608(3) &  0.02599(2)    \\
$160$  & 0.28001(10)   &  0.09181(5) &   0.03333(3) &  0.01302(1)     \\
$320$  & 0.23505(9)    &  0.06477(4) &   0.01977(2) &  0.006503(5)   
\end{tabular}
\end{center}
\label{cpmom}
\end{table}
 
\begin{table}
\caption{\sf Cumulants and ratios for the critical CP.}
\begin{center}
\begin{tabular}{|r|l|l|l|l|} 
$L$   &  $K_2$  & $K_2/m_1^2 $ &  $K_4/K_2^2$  &  $m_4/m_2^2 $  \\
\hline\hline
$20$  &   0.03507(12)  &  0.1559(12)  & -0.641(4) & 1.490(1) \\
$40$  &   0.02596(1)   &  0.1643(1)   & -0.577(2) & 1.520(2) \\
$80$  &   0.01877(1)   &  0.1686(2)   & -0.544(3) & 1.535(2) \\
$160$ &   0.01341(1)   &  0.1710(2)   & -0.525(3) & 1.545(3) \\
$320$ &   0.009517(3)  &  0.1723(3)   & -0.515(4) & 1.550(2)      
\end{tabular}
\end{center}
\label{cprat}
\end{table}

\begin{table}
\caption{\sf Density moments for the critical A model.}
\begin{center}
\begin{tabular}{|c|c|c|c|c|} 
$L$    &    $m_1$ &       $ m_2 $    &    $m_3$     &  $m_4$ \\
\hline\hline
$20$  &   0.43792(10)    &  0.22144(8)   &   0.12305(5)      &   
0.07316(3)    \\
$40$  &   0.36672(3)     &  0.15648(2)   &   0.07366(2)      &   
0.03721(1)    \\
$80$  &   0.30770(3)     &  0.11059(2)   &   0.04395(2)      &   
0.01878(2)    \\
$160$  & 0.25812(5)      &  0.07800(3)    &  0.02609(1)      &   
0.009398(5)     \\
$320$  & 0.21691(4)      &  0.05513(3)    &  0.01552(1)      &   
0.004707(4)   
\end{tabular}
\end{center}
\label{ammom}
\end{table}
 
\begin{table}
\caption{\sf Cumulants and ratios for the critical A model.}
\begin{center}
\begin{tabular}{|r|l|l|l|l|} 
$L$   &  $K_2$  & $K_2/m_1^2 $ &  $K_4/K_2^2$  &  $m_4/m_2^2 $  \\
\hline\hline
$20$  &  0.02972(6)  &  0.1550(4) &  -0.608(4) & 1.492(2)       \\
$40$  &  0.021995(4) &  0.1636(1) &  -0.566(1) & 1.520(1)      \\
$80$  &  0.015914(4) &  0.1681(1) &  -0.538(1) & 1.536(1)     \\
$160$  & 0.011375(5) &  0.1707(1) &  -0.522(2) & 1.545(2)      \\
$320$  & 0.008082(5) &  0.1718(2) &  -0.512(8) & 1.549(3)      
\end{tabular}
\end{center}
\label{amrat}
\end{table}

\begin{table}
\caption{\sf Density moments for the critical PCP.}
\begin{center}
\begin{tabular}{|c|c|c|c|c|} 
$L$    &    $m_1$ &       $ m_2 $    &    $m_3$     &  $m_4$ \\
\hline\hline
$20$  &   0.51654(8) &  0.31178(4)   &   0.20815(4) &   0.14927(4)    \\
$40$  &   0.43101(3) &  0.21800(2)   &   0.12228(2) &   0.07390(1)    \\
$80$  &   0.36123(3) &  0.15327(2)   &   0.07215(1) &   0.03663(1)    \\
$160$  & 0.30308(4)  &  0.10789(1)   &   0.04262(1) &   0.01815(1)     \\
$320$  & 0.25448(4)  &  0.07604(1)   &   0.02521(1) &   0.009010(3)   \\
$640$  & 0.21373(10) &  0.05363(3)   &   0.01493(2) &   0.004480(5)   
\end{tabular}
\end{center}
\label{pcpmom}
\end{table}
 
\begin{table}
\caption{\sf Cumulants and ratios for the critical PCP.}
\begin{center}
\begin{tabular}{|r|l|l|l|l|} 
$L$   &  $K_2$  & $K_2/m_1^2 $ &  $K_4/K_2^2$  &  $m_4/m_2^2 $  \\
\hline\hline
$20$  & 0.044921(3)  &  0.16836(6) &   -0.66102(3) & 1.5356(8)  \\
$40$  & 0.032230(4)  &  0.17349(4) &   -0.55536(3) & 1.5550(5)  \\
$80$  & 0.022779(5)  &  0.17457(7) &   -0.5152(5)  & 1.5593(8)  \\
$160$ & 0.016038(6)  &  0.1746(1)  &   -0.5014(4)  & 1.559(1) \\
$320$ & 0.011279(5)  &  0.1742(1)  &   -0.496(2)   & 1.558(1)   \\
$640$ & 0.007949(5)  &  0.1740(3)  &   -0.497(3)   & 1.558(4)     
\end{tabular}
\end{center}
\label{pcprat}
\end{table}

In Fig. 3 we plot the ratios $K_2/m_1^2 $ versus $L^{-1}$
for each of the three
one-dimensional models.  For the CP and the A model we
observe a very similar, nearly linear approach to a limit.  (Least-squares
quadratic fits of $K_2/m_1^2 $ versus $L^{-1}$
yield limiting values of 0.1736 for the CP, 0.1732 for the A model,
and 0.1738 for the PCP.)
The PCP ratio initially approaches its limit more rapidly
than the CP and A model do, overshoots the limit, and, for
large $L$, approaches the limit from above.
Thus it would appear that the
dominant correction to scaling for the size-dependence
of $P(\rho,L)$ is
different in the PCP than in the CP and the A model.
We conclude that for the DP class in 1+1 dimensions,
$K_2/m_1^2 = 0.1735(5)$, with the uncertainty figure a subjective
assessment based on the degree of regularity of the data.

The ratios $m_4 /m_2^2 $
in the three one-dimensional models are plotted
versus $L^{-1}$ in Fig. 4.  As before, the CP and
A model exhibit very similar
trends, and the PCP has a nonmonotonic
approach to its apparent limit.
Quadratic fits yield limiting values of
1.554, for the CP and the A model, and 1.558 for PCP.
This suggests a value of 1.556(3) for the DP class
in 1+1 dimensions.   Our estimates for the cumulant ratio,
$K_4/K_2^2$, are considerably more scattered.
Quadratic fits to the data for
for $L \geq 40$ yield limiting values of
-0.505, -0.503, and -0.493 for the CP, A model, and PCP,
respectively.  We estimate the value of this ratio as -0.50(1) for 
the DP class in 1+1 dimensions.

\vspace{1em}

\section{Contact Process in Two Dimensions}

In this section we report results for the two 
dimensional CP at the critical point, 
$\lambda = \lambda_c=1.6488$\cite{agrd96}.  We studied systems
of $L \times L$ sites with $L$ ranging from 10 to 160,
with a sample size ranging from $3 \times 10^6$ for $L=10$
to $10^6$ for $L=160$.
The maximum duration of a trial runs 
from $1000$ for $L=10$ to $10^5$ for $L=160$.
We calculate moments from a sample of
$10^4$ maximum-duration trials for $L=10$ to $L=80$,
and of $5000$ such trials for the largest $L$. 
Density moments and ratios are
given in Tables
\ref{cp2mom} and \ref{cp2rat}, respectively.  The results indicate that
$K_2/m_1^2 = 0.3264(5)$ for the DP class in 2+1
dimensions.  As in one dimension, $m_4/m_2^2$ is particularly
stable, and here takes the value $2.094(8)$.
Our results for $K_4/K_2^2$ are less precise;
we conclude that this ratio takes a value of $-0.088(4)$.

\begin{table}
\caption{\sf Density moments for the critical CP in two dimensions.}
\begin{center}
\begin{tabular}{|r|l|l|l|l|} 
$L$   &    $m_1$    &   $ m_2 $     &    $m_3$       &  $m_4$ \\
\hline\hline
$10$  & 0.15706(6)  &  0.03228(2)   &   0.00781(1)   &   0.002120(3)    \\
$20$  & 0.09027(6)  &  0.01078(1)   &   0.001522(2)  &   0.0002416(4)    \\
$40$  & 0.05200(3)  &  0.003587(3)   &   0.0002929(3)  &   0.00002693(4) \\
$80$  & 0.03003(2)  &  0.001197(2)   &   0.0000564(1)  &   0.00000300(1)\\
$160$ & 0.01736(1)  &  0.0003998(5)  &  0.00001090(2)  & $0.335(1)\times 10^{-6}$  
\end{tabular}
\end{center}
\label{cp2mom}
\end{table}
 
\begin{table}
\caption{\sf Cumulants and ratios for the critical CP in two dimensions.}
\begin{center}
\begin{tabular}{|r|l|l|l|l|} 
$L$   &  $K_2$      & $K_2/m_1^2$ & $K_4/K_2^2$ &  $m_4/m_2^2 $  \\
\hline\hline
$10$  & 0.00761(1)   &  0.3085(3)  & -0.169(3)   & 2.034(6)    \\
$20$  & 0.002631(2)  &  0.3229(4)  & -0.115(3)   & 2.079(8)    \\
$40$  & 0.0008832(7) &  0.3267(4)  & -0.090(4)   & 2.093(7)    \\
$80$  & 0.0002946(3) &  0.3269(4)  & -0.089(5)   & 2.095(8)\\
$160$ & 0.0000983(1) &  0.3264(5)  & -0.088(4)   & 2.094(8) \\       
\end{tabular}
\end{center}
\label{cp2rat}
\end{table}

\section{Summary}

We have determined order parameter moments and cumulant ratios
for one and two dimensional models in the directed percolation
universality class.  In one dimension, we studied three different
models: the contact process, the closely-related A model, and
the pair contact process.  Our results lead to the estimates:
$K_2/m_1^2 = 0.1735(5)$,
$m_4 /m_2^2 = 1.556(3)$
and $K_4/K_2^2 = -0.50(1)$,
for the DP class in 1+1 dimensions.  We also derived an
improved estimate, $p_c = 0.077090(5)$, for the critical point of the
one-dimensional pair contact process.  In two dimensions we
restricted our attention to the basic contact process, obtaining
$K_2/m_1^2 = 0.3264(5)$, $m_4/m_2^2 = 2.094(8)$,
and $K_4/K_2^2 = -0.088(4)$.

\newpage
\noindent {\bf Figure Captions} 
\vspace{1em}

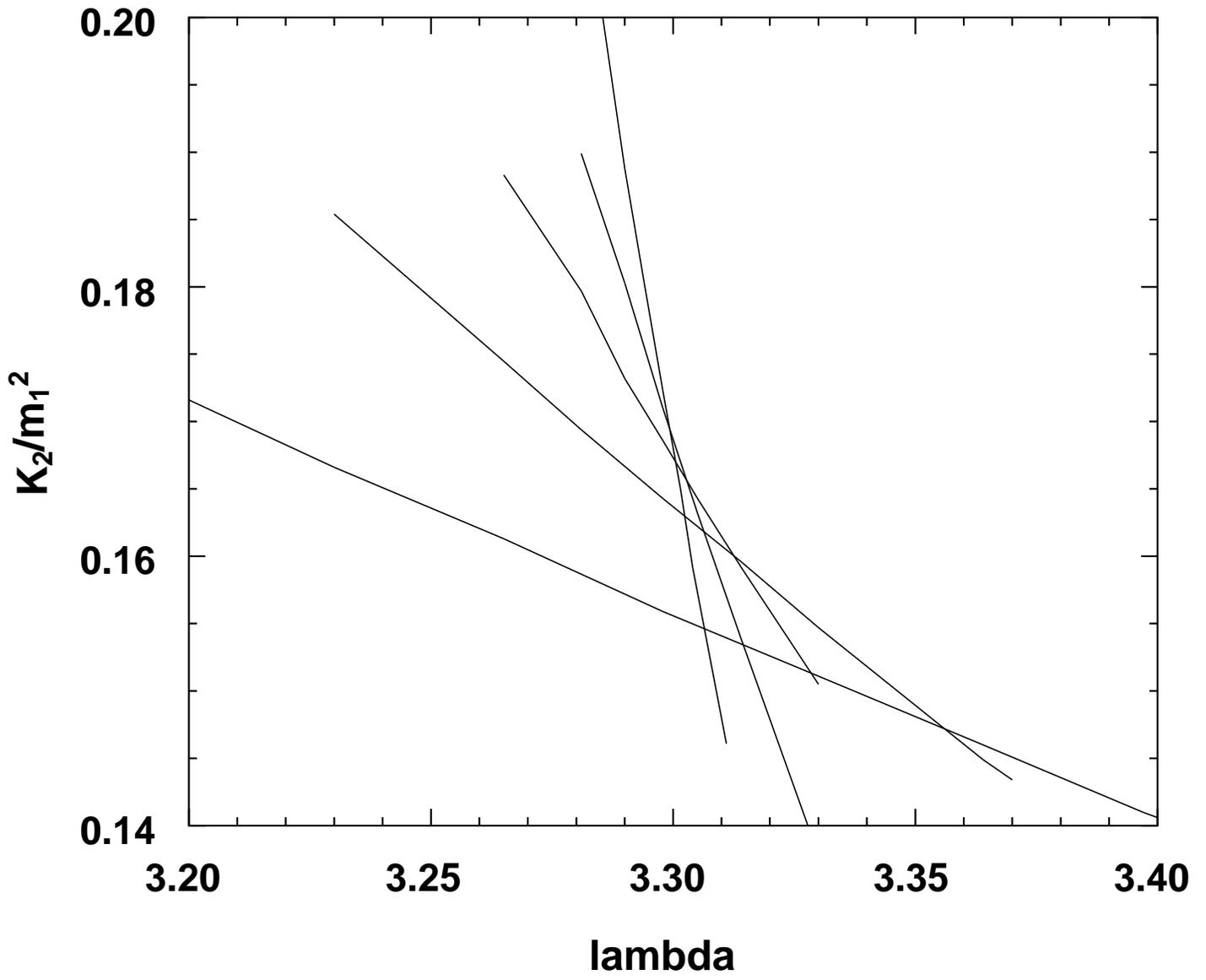
\noindent FIG. 1. The ratio $K_2/m_1^2 $ versus
creation rate $\lambda$ in the one-dimensional CP.
System sizes $L=20, 40, ...,320$, in order of
increasing steepness.
\vspace{1em}

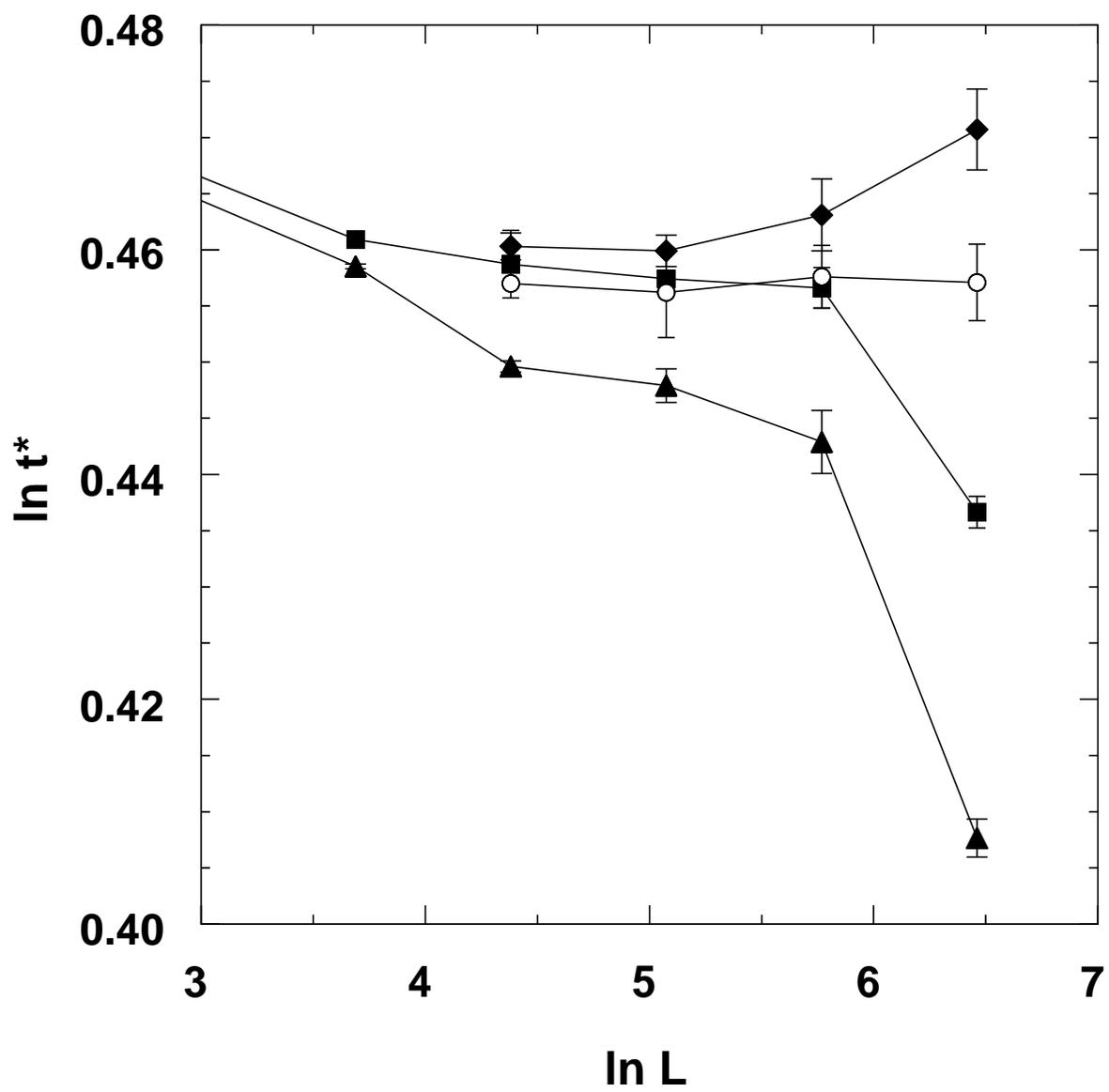
\noindent FIG. 2. Scaled lifetime
$t^* \equiv \tau/L^{\nu_{||}/\nu_{\perp}} $
versus system size $L$ in the PCP.
Diamonds: $p=0.07708$; $\circ$: 0.07709;
squares: 0.07710; triangles: 0.07712.
\vspace{1em}

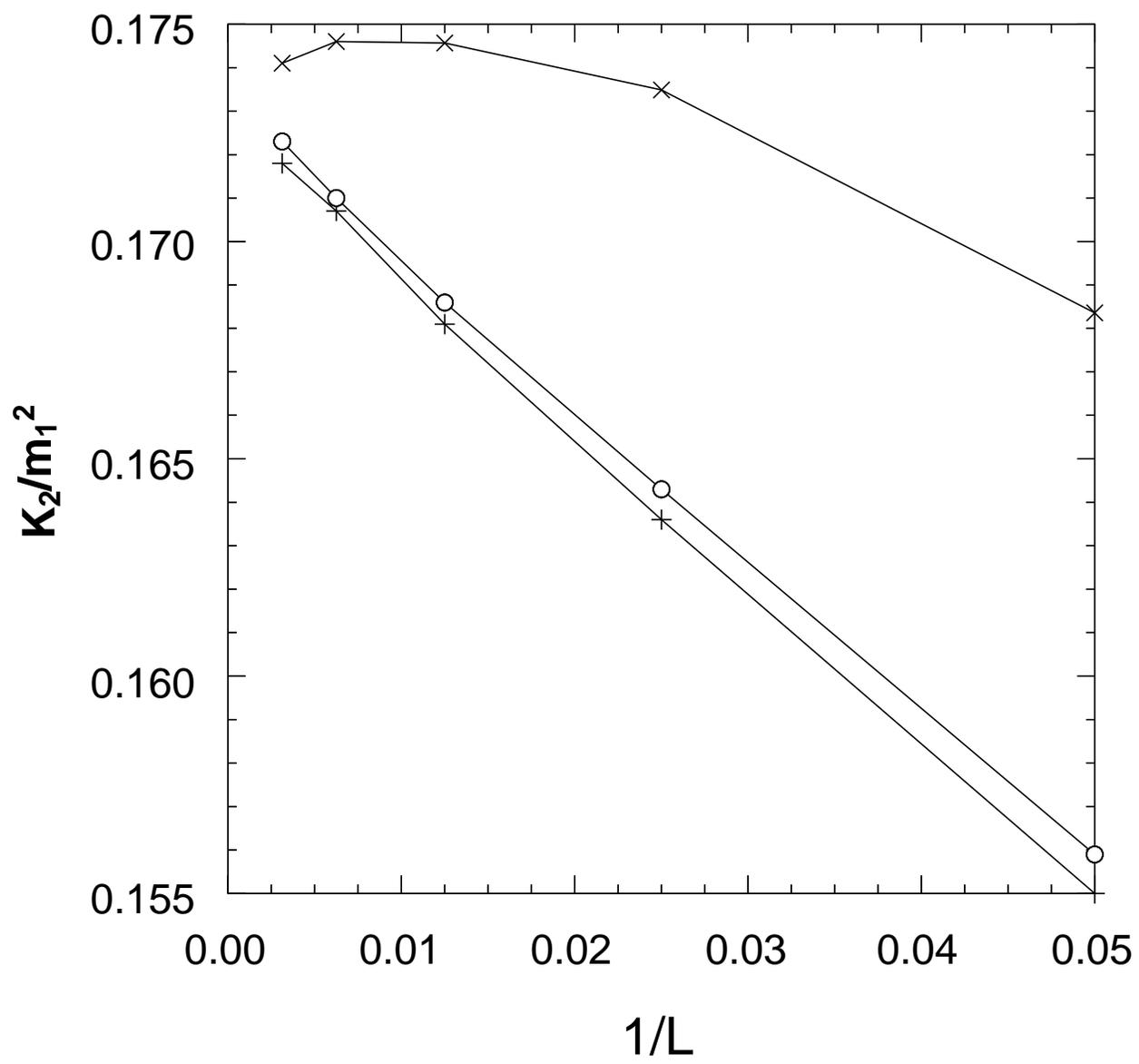
\noindent FIG. 3. $K_2/m_1^2 $ versus $1/L$ for one-dimensional
models at their critical points.  $\circ$: CP; +: A model; $\times$: PCP.
\vspace{1em}

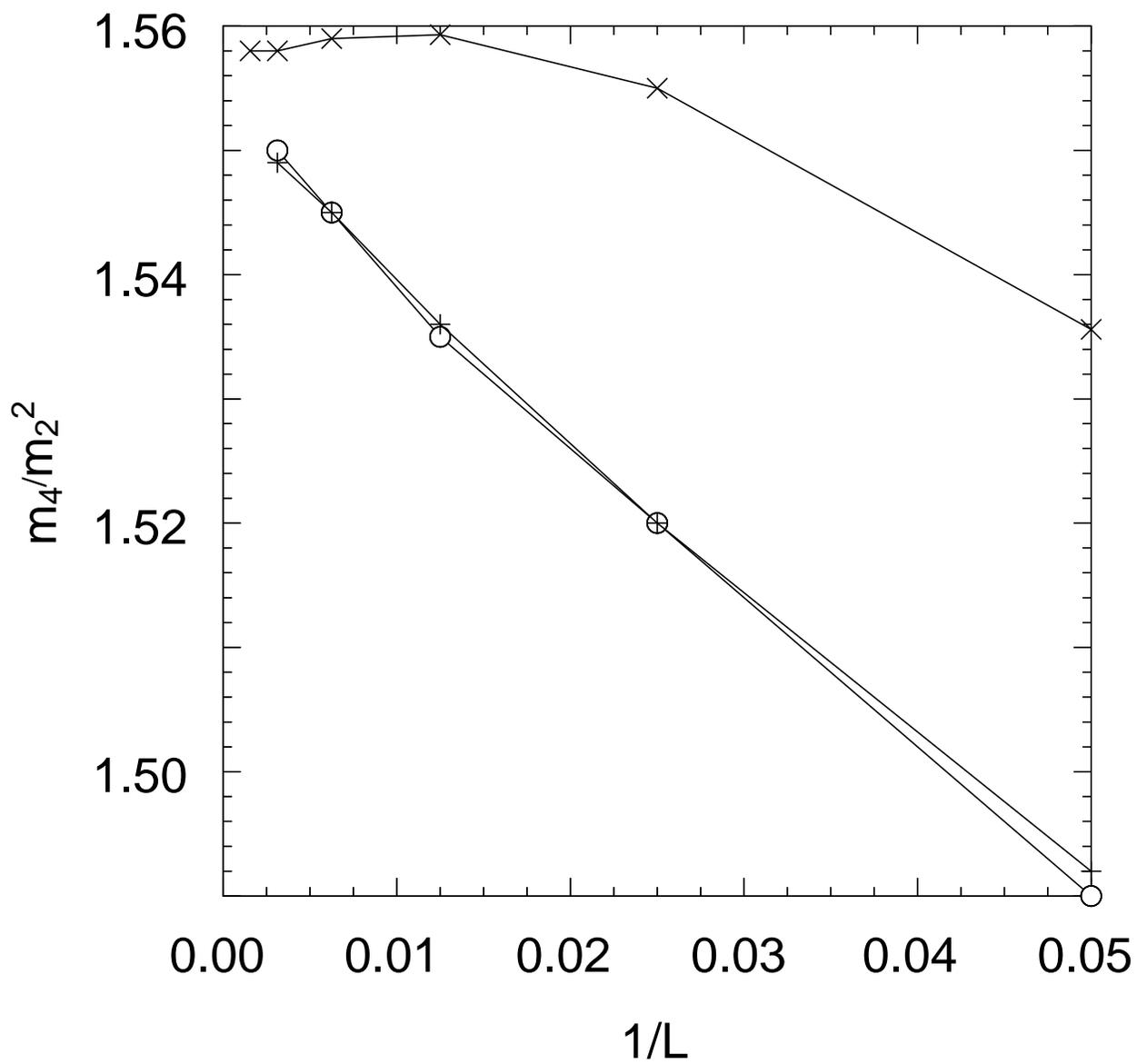
\noindent FIG. 4. Same as Fig. 3, but for $m_4/m_2^2$.

\end{document}